# NETWORK TRAFFIC ANALYSIS: HADOOP PIG VS TYPICAL MAPREDUCE

Anjali P P[1] and Binu A[2]

[1]Department of Information Technology, Rajagiri School of Engineering and Technology, Kochi. M G University, Kerala
anjalinambiarpp@gmail.com

[2]Department of Information Technology, Rajagiri School of Engineering and Technology, Kochi. M G University, Kerala
binu_a@rajagiritech.ac.in

## ABSTRACT

*Big data analysis has become much popular in the present day scenario and the manipulation of big data has gained the keen attention of researchers in the field of data analytics. Analysis of big data is currently considered as an integral part of many computational and statistical departments. As a result, novel approaches in data analysis are evolving on a daily basis. Thousands of transaction requests are handled and processed everyday by different websites associated with e-commerce, e-banking, e-shopping carts etc. The network traffic and weblog analysis comes to play a crucial role in such situations where Hadoop can be suggested as an efficient solution for processing the Netflow data collected from switches as well as website access-logs during fixed intervals.*

## 1. INTRODUCTION

The network protocol designed by Cisco Systems for assembling network traffic information named as NetFlow has become an industry standard for monitoring traffic and is supported by various platforms. Cisco standard NetFlow version 5 defines network flow as a unidirectional sequence of packets which has got seven parameters:

- Ingress interface (SNMP ifIndex)
- Source IP address
- Destination IP address
- IP protocol
- Source port for UDP or TCP, 0 for other protocols
- Destination port for UDP or TCP, type and code for ICMP, or 0 for other protocols
- IP Type of Service

For limiting the load on the router components and the amount of flow records exported, Netflow is normally enabled on a per-interface basis. If IP filters are not used to decide if a packet can be captured by NetFlow, it will capture all packets received by an ingress interface by default. Further customizations can also be done on NetFlow implementation as per the requirement in different scenarios. It can even allow the observation and recording of IP packets on the egress interface if accurately configured.

NetFlow data from dedicated probes can be efficiently utilized for the observation of critical links, whereas the data from routers can provide a network-wide view of the traffic that which will be of great use for capacity planning, accounting, performance monitoring, and security.

This paper depicts a survey regarding the network flow data analysis done using Apache Hadoop MapReduce framework and a comparative study of typical MapReduce against Apache Pig, a platform for analyzing large data sets and provides ease of programming, optimizing opportunities and enhanced extensibility.

## 2. NETFLOW ANALYSIS USING TYPICAL HADOOP MAPREDUCE FRAMEWORK

Hadoop can be considered as an efficient mechanism for big data manipulation in distributed environment. Hadoop was developed by Apache foundation and is well-known for its scalability and efficiency. Hadoop works on a computational paradigm called MapReduce which is characterized by a Map and a Reduce phase. In Map phase, the computation is divided into many fragments and is distributed among the cluster nodes. The individual results evolved at the end of the Map phase is combined and reduced to a single output in the Reduce phase, thereby producing the result in a short span of time.

Hadoop will store the data in its distributed file system popularly known as HDFS and will make it easily available to the users when requested. HDFS is greatly advantageous in its scalability and portability aspects. It achieves reliability by the replication of data across multiple nodes in the cluster, thereby eliminating the need for RAID storage on the cluster nodes. High availability is also incorporated in Hadoop cluster systems by means of a backup provision for the master node[name node] known as secondary namenode. The major limitations of HDFS are the inability to be mounted directly by an existing operating system and the inconvenience of moving data in and out of the HDFS before and after the execution of a job.

Once Hadoop is deployed on a cluster, instances of these processes will be started – Namenode, Datanode, Jobtracker, Tasktracker. Hadoop can process large amount of data such as website logs, network traffic logs obtained from switches etc. The namenode, otherwise known as master node is the main component of the HDFS which stores the directory tree of the file system and keeps track of the data during execution. The datanode stores the data in HDFS as suggested by its name. The jobtracker delegates the MapReduce tasks to different cluster nodes and the tasktracker spawns and executes the job on the member nodes.

The Netflow analysis requires sufficient network flow data collected at a particular time or at regular intervals for processing. Upon analyzing the incoming and outgoing network traffic corresponding to various hosts, required precautions can be taken regarding server security, firewall setup, packet filtering etc and here, the significance of network flow analysis cannot be kept aside. Flow data can be collected from switches or routers using router-console commands and are filtered on the basis of the required parameters such as IP address, Port numbers etc. Similarly, every website will generate corresponding logs and such files can be collected from the web-server for further analysis. Website access logs will contain the details of URLs accessed, the time of access, number of hits etc which are unavoidable for a website manager for the proactive detection and encounter of problems like Denial of Service attacks and IP address spoofing.

The log files generated by a router or web-server every minute will accumulate to form a large-sized file which needs a great deal of time to get processed. In such a situation, the advantage of Hadoop MapReduce paradigm can be made use of. The big data is fed as input for a MapReduce algorithm which provides an efficiently processed output file in return. Every map function transforms a data into certain number of key/value pairs which later sorts them according to the key values whereas a reduce function merges the values with similar key values into a single result.

One of the main feature of Hadoop MapReduce is its coding complexity which is possible only by programmers with highly developed skills. This programming model is highly restrictive and Hadoop cluster management is also not much easy while considering aspects like debugging, log collection, software distribution etc. In short, the effort to be put in MapReduce programming restricts its acceptability to a remarkable extent. Here comes the need for a platform which can provide ease of programming as well as enhanced extensibility and Apache Pig reveals itself as a novel approach in data analytics. The key features and applications of Apache Pig deployed over Hadoop framework is described in Section 3.

## 3. USING APACHE PIG TO PROCESS NETWORK FLOW DATA

Apache Pig is a platform that supports substantial parallelization which enables the processing of huge data sets. The structure of Pig can be described in two layers – An infrastructure layer with an in-built compiler that can produce MapReduce programs and a language layer with a text-processing language called "Pig Latin". Pig makes data analysis and programming easier and understandable to novices in Hadoop framework using Pig Latin which can be considered as a parallel data flow language. The pig setup has also got provisions for further optimizations and user defined functionalities. Pig Latin queries are compiled into MapReduce jobs and are executed in distributed Hadoop cluster environments. Pig Latin looks similar to SQL[Structured Query Language] but is designed for Hadoop data processing environment just like SQL for RDBMS environment. Another member of Hadoop ecosystem known as Hive is a query language like SQL and needs data to be loaded into tables before processing. Pig Latin can work on schema-less or inconsistent environments and can operate on the available data as soon as it is loaded into the HDFS. Pig closely resembles scripting languages like Perl and Python in certain aspects such as the flexibility in syntax and dynamic variables definitions. Pig Latin is efficient enough to be considered as the native parallel processing language for distributed systems such as Hadoop.

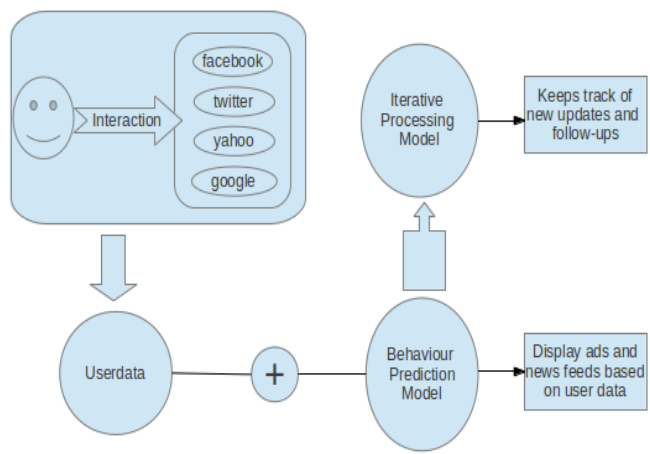

Figure 1. Data processing model in PIG

Pig is gaining popularity in different zones of computational and statistical departments. It is widely used in processing weblogs and wiping off corrupted data from records. Pig can build behavior prediction models based on the user interactions with a website and this feature can be applied in displaying ads and news stories that keeps the user entertained. It also emphasizes on an iterative processing model which can keep track of every new updates by combining the behavioral model with the user data. This feature is largely used in social networking sites like

Facebook and micro-blogging sites like Twitter. When it comes to small data or scanning multiple records in random order, pig cannot be considered as effective as MapReduce. The aforementioned data processing model used by Pig is depicted in Figure 1.

In this paper, we are exploring the horizons of Pig Latin for processing the netflow obtained from routers. The proposed method for implementing Pig for processing netflow data is described in Section 4.

## 4. IMPLEMENTATION ASPECTS

Here we present the NetFlow analysis using Hadoop, which can manage large volume of data, employ parallel processing and come up with required output in no time. A map-reduce algorithm need to be deployed using Hadoop to get the result and writing such map-reduce programs for analyzing huge flow data is a time consuming task. Here the Apache-Pig will help us to save a good deal of time.

We need to analyze traffic at a particular time and Pig plays an important role in this regard. For data analysis, we create buckets with pairs of (SrcIF, SrcIPaddress), (DstIF,DstIPaddress), (Protocal,FlowperSec). Later, they are joined and categorized as SrcIF and FlowperSec; SrcIPaddress and FlowperSec. Netflow data is collected from a busy Cisco switch using nfdump. The output contains three basic sections-IP Packet section, Protocol Flow section and Source Destination packets section.

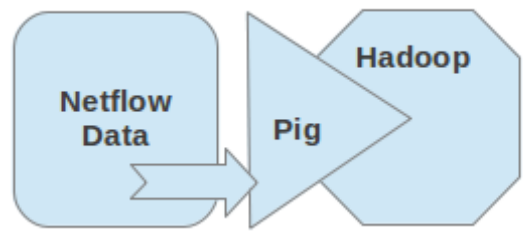

Figure 2. Netflow Analysis using Hadoop Pig

Pig needs the netflow datasets in arranged format which is then converted to SQL-like schema. Hence the three sections are created as three different schemas. A sample Pig Latin code is given below:

Protocol = LOAD 'NetFlow-Data1' AS (protocol:chararray, flow:int, …)
Source = LOAD 'NetFlow-Data2' AS (SrcIF:chararray, SrcIPAddress:chararray, …)
Destination = LOAD 'NetFlow-Data3' AS (DstIF:chararray, DstIPaddress, …)

As shown in Figure 2, the network flow is deduced into schemas by Pig. Then these schemas are combined for analysis and the traffic flow per node is the resultant output.

## 5. ADVANTAGES OF USING HADOOP PIG OVER MAPREDUCE

The member of Hadoop ecosystem popularly known as Pig can be considered as a procedural version of SQL. The most attractive feature and advantage of Pig is its simplicity. It doesn't demand highly skilled Java professionals for implementation. The flexibility in syntax and dynamic variable allocation makes it much more user-friendly and provisions for user-defined functions add extensibility for the framework. The ability of Pig to work with un-structured data greatly accounts to parallel data processing. The great difference comes when the computational complexity is considered. Typical map-reduce programs will have large number of java code lines whereas the task can be accomplished using Pig Latin in very fewer lines of code. Huge network flow data input files can be processed with writing simple programs in Pig Latin which closely resembles scripting languages like Perl/Python. In other words, the computational complexity of Pig Latin is lesser than Hadoop MapReduce programs written in Java which made Pig suitable for the implementation of this project. The procedure followed for arriving at such a conclusion is described in Section 6.

## 6. EXPERIMENTS AND RESULTS

Hadoop Pig was selected as the programming language for Netflow analysis after conducting a short test on a sample input file.

### 6.1 Experimental Setup

Apache Hadoop was initially installed and configured on a stand-alone node and Pig was deployed on top of this system. Key-based authentication was setup on the machine as Hadoop prompts for passwords while performing operations. Hadoop instances such as namenode, datanode, jobtracker and tasktracker are started before running Pig and it shows a "grunt" console upon starting. The grunt console is the command prompt specific to Pig and is used for data and file system manipulation. Pig can work both in local mode and MapReduce mode which can be selected as per the user requirement.

A sample Hadoop MapReduce word-count program written in Java was used for the comparison. The program used to process a sample input file of enough size had one hundred thirty lines of code and the time taken for execution was noted. The same operation was coded using Pig Latin which contained only five lines of code and corresponding running time was also noted for framing the results.

### 6.2. Experiment Results

At a particular instant, the size of input file is kept constant for MapReduce and Pig whereas the lines of code are more for the former and less for the latter. For a good comparison, the input file size was taken in the range of 13 kb to 208 kb and the corresponding time is recorded as in Figure 3.

| Turn | Operation | Input File Size(kb) | Time Taken by MapReduce (Lines of Code : 130) | Time Taken by Pig (Lines of Code : 5) |
|---|---|---|---|---|
| 1 | Word-Count | 13 kb | 34 sec | 31 sec |
| 2 | Word-Count | 26 kb | 33 sec | 31 sec |
| 3 | Word-Count | 52 kb | 35 sec | 31 sec |
| 4 | Word-Count | 104 kb | 35 sec | 31 sec |
| 5 | Word-Count | 208 kb | 36 sec | 31 sec |

Figure 3. Hadoop Pig vs. MapReduce

From this we can formulate an expression as follows: As the input file size increases in the multiples of x, the execution time for typical MapReduce also increases proportionally whereas Hadoop Pig can maintain a constant time at least for x upto 5 times. Graphs are plotted between both and are given as Figure 4 and Figure 5.

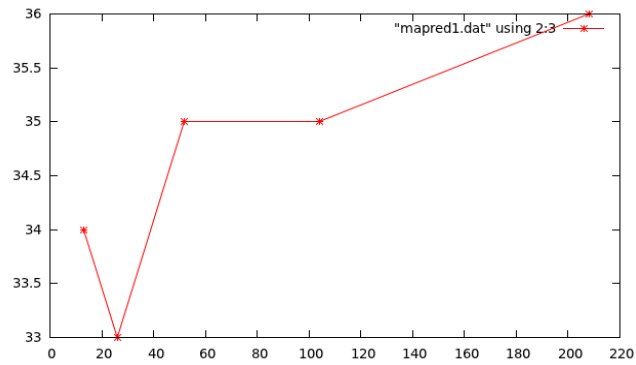

Figure 4. Input file-size vs. execution time for MapReduce

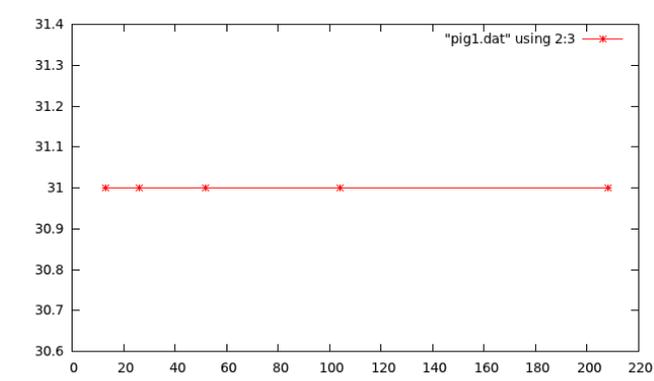

Figure 5. Input file-size vs. execution time for Pig

## 6. CONCLUSION

In this paper, a detailed survey regarding the wide scope of Apache Pig platform is conducted. The key features of the network flows are extracted for data analysis using Hadoop Pig which was tested and proved to be advantageous in this aspect with a very low computational complexity. Network flow capture is to be done with the help of easily available open-source tools. Once the Pig-based flow analyzer is implemented, traffic flow analysis can be done with increased convenience and easiness even without the help of highly developed programming skills. It greatly reduces the time consumed for researching on the complex programming constructs and control loops for implementing a typical MapReduce paradigm using Java, but still enjoying the advantages of Map/Reduce task division method with the help of the layered architecture and in-built compilers of Apache Pig.

## ACKNOWLEDGEMENT

We are greatly indebted to the college management and the faculty members for providing necessary facilities and hardware along with timely guidance and suggestions for implementing this work.

**Authors**

**Anjali P P** is a MTech student of Information Technology Department in Rajagiri college of engineering, Kochi, under MG University. She has received BTech degree in Computer Science and Engineering from Cochin University of Science and Technology. Her research interest includes Hadoop and Big data analysis.

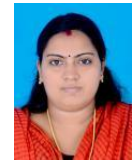

**Binu A** is an Assistant Professor in Information Technology Department of Rajagiri college of engineering, Kochi, under MG University.

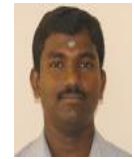